\documentclass[12pt,preprint]{aastex}

\shortauthors{Matthews \& Uson}
\shorttitle{Disk Corrugations in IC~2233}

\begin{document}

\newcommand{\ang}{\rm \AA}
\newcommand{\msun}{M$_\odot$}
\newcommand{\lsun}{L$_\odot$}
\newcommand{\days}{$d$}
\newcommand{\degree}{$^\circ$}
\newcommand{\ud}{{\rm d}}
\newcommand{\as}[2]{$#1''\,\hspace{-1.7mm}.\hspace{.0mm}#2$}
\newcommand{\am}[2]{$#1'\,\hspace{-1.7mm}.\hspace{.0mm}#2$}
\newcommand{\ad}[2]{$#1^{\circ}\,\hspace{-1.7mm}.\hspace{.0mm}#2$}
\newcommand{\lsim}{~\rlap{$<$}{\lower 1.0ex\hbox{$\sim$}}}
\newcommand{\gsim}{~\rlap{$>$}{\lower 1.0ex\hbox{$\sim$}}}
\newcommand{\HA}{H$\alpha$}
\newcommand{\HII}{\mbox{H\,{\sc ii}}}
\newcommand{\NII}{\mbox{N\,{\sc ii}}}
\newcommand{\kms}{\mbox{km s$^{-1}$}}
\newcommand{\HI}{\mbox{H\,{\sc i}}}
\newcommand{\KI}{\mbox{K\,{\sc i}}}
\newcommand{\popI}{\mbox{Population\,{\sc i}}}
\newcommand{\nan}{Nan\c{c}ay}
\newcommand{\galex}{{\it GALEX}}
\newcommand{\jks}{Jy~km~s$^{-1}$}

\title{Corrugations in the Disk of the Edge-On Spiral
Galaxy IC~2233}
\author{L. D. Matthews\altaffilmark{1} \& Juan M. Uson\altaffilmark{2}}

\altaffiltext{1}{Harvard-Smithsonian Center for Astrophysics,
60 Garden Street, Cambridge, MA, USA 02138}
\altaffiltext{2}{National Radio Astronomy Observatory, 520 Edgemont
Road, Charlottesville, VA USA 22903}

\begin{abstract}
We recently reported the discovery of a
regular corrugation pattern in the \HI\ disk of
the isolated, edge-on spiral galaxy IC~2233. Here we present
measurements of the vertical structure of this galaxy at several
additional wavelengths, ranging from the far ultraviolet to the far 
infrared. We find that undular patterns with amplitude $\lsim 5''$
($\lsim$250~pc) are 
visible in a variety of \popI\ tracers in IC~2233, including the
young-to-intermediate age stars, the \HII\ regions, and the dust. 
However, the 
vertical excursions become less pronounced in the older stellar populations 
traced by the
mid-infrared light. This suggests that the process leading to the
vertical displacements may be linked with the regulation of star
formation in the galaxy. We have also identified 
a relationship between the locations
of the density corrugations and small-amplitude ($\lsim$5~\kms) 
velocity undulations in the \HI\
rotation curve. We are able to exclude several possible mechanisms
for the origin of the observed corrugations, including tidal
interaction from a companion, Parker instabilities, or a galactic bore.  
Global gravitational instabilities appear to be the most 
likely explanation, although 
local perturbations may also be important.
\end{abstract}

\keywords{galaxies: spiral -- galaxies: Individual (IC~2233) --
galaxies: structure -- ISM: structure -- instabilities}

\section{Introduction}
It is well known that the disks of late-type galaxies are 
not completely 
flat. Warping of the outer gas disk into an 
``integral sign''
shape is a nearly ubiquitous feature of
spiral galaxies, even those that appear to be quite isolated
(Garc\'\i a-Ruiz et al. 2002). A significant fraction of
galaxies also have warped stellar disks 
(S\'anchez-Saavedra et al. 1990; Reshetnikov \& Combes 1998).
For several decades it has been recognized that the \HI\ disk and other
\popI\ components of the Milky Way
exhibit an additional type of vertical structure in the form of
systematic displacements from the mean plane known as
``corrugations''
(e.g., Gum et al. 1960; Quiroga 1974,1977; Lockman 1977; Sanders et
al. 1984). These
corrugations are present along both radial and azimuthal directions
and have scales ranging from $\sim$50 to 350~pc in amplitude and from
one to several kiloparsecs in wavelength (Spicker \& Feitzinger
1986). 

Despite
predictions that corrugations may be a common phenomenon in disk galaxies,
systematic searches for these features have been limited, and
consequently, little has
been known about their true frequency of occurrence in external galaxies. 
To date, evidence of corrugations has been reported 
for only a handful of cases outside the Milky Way: in 
the emission nebulae of M31 (Arp 1964), in the 
stellar (optical) light of three edge-on, late-type
galaxies (Florido et al. 1991,1992), and in the velocity structure of
the face-on spiral NGC~5427 (Alfaro et al. 2001). 

The origin of disk corrugations is a matter of longstanding debate
given the small number of galaxies in which they have so far been detected
(see the reviews by 
Alfaro \& Efremov 1996 and  
Alfaro 2003). Suggestions have included:
gravitational instabilities (e.g., Nelson 1976); tidal interactions (Edelsohn
\& Elmegreen 1997), collisions of high-velocity clouds with the disk
(Franco et al. 1988; 
Santill\'an et al. 1999); interaction of spiral waves with the gaseous
disk (Alfaro et al. 2001); and the undular mode of the Parker
instability (Franco et al. 2002).
Depending on their origin, the presence or absence of corrugations 
can provide important
clues into the role of magnetic fields on galaxy disk
structure (e.g., Spicker \&
Feitzinger 1986; Santill\'an et al. 2000; Franco et al. 2002), 
the formation of
molecular clouds (e.g., Nelson \& Matsuda 1980),
the degree of self-gravity of galaxy disks (e.g.,
Revaz \& Pfenniger 2004), or the timescales of recent interactions
(e.g., Edelsohn \& Elmegreen 1997). Moreover,
current evidence suggests that whatever the physical 
process(es) responsible for corrugations, they are
closely linked to the mechanisms responsible
for the formation of the dense gas condensations and therefore with the
regulation of star formation 
(e.g., Nelson \& Matsuda 1980; Alfaro et al. 1992; 
Alfaro \& Efremov 1996; Alfaro 2003). 

In a recent paper (Matthews \& Uson 2008; hereafter MU08), 
we presented  optical and \HI\
imaging observations of the nearby, edge-on spiral galaxy IC~2233. 
There we reported the discovery of a 
corrugated pattern in the \HI\ disk of this galaxy---i.e., we found that 
the \HI\ layer of IC~2233 exhibits a remarkably regular pattern of
positive and negative vertical displacements from the kinematically-defined 
midplane. This pattern has a wavelength of $\sim150''$
(7~kpc) and an amplitude that increases with distance from the center
of the galaxy, reaching a maximum of $\sim3''$
($\sim$150~pc).\footnote[3]{Throughout this paper we assume a distance to
IC~2233 of 10~Mpc (see MU08).} [For comparison, the
FWHM thickness of the \HI\ layer ranges from $\sim$\as{10}{4} near the
disk center to $\sim$\as{20}{3} toward the outer disk (MU08)]. 
Subsequently, we have explored the vertical
structure of IC~2233 using a variety of additional tracers of the
stellar and interstellar components of its disk. We have also examined
the relationship between the peaks of the vertical \HI\ excursions
with both \HI\ density enhancements along the disk and with 
velocity perturbations in the disk rotation curve. Here
we report the discovery of undular patterns in several additional
\popI\ tracers in IC~2233. 
In contrast, we find the vertical
displacements are minimal in the old stars. We also report the discovery of a
link between the locations of the peak excursions from the
midplane in \HI\ with features in the \HI\ rotation curve.
In \S~\ref{discussion} we  discuss the implications of these findings
for understanding the origin of the disk corrugations seen in
IC~2233 and other galaxies.

\section{A Multiwavelength Analysis of the Vertical Structure of
IC~2233}
IC~2233 is a bulgeless Sd galaxy located at a
distance of $\sim$10~Mpc. It has a moderately low surface brightness, 
``superthin'' disk structure (Goad
\& Roberts 1981) and a low current star formation rate 
($\lsim0.05~M_{\odot}$~yr$^{-1}$; MU08).  
The transparency and structural simplicity 
of IC~2233 make it well-suited to a detailed
vertical structure analysis. IC~2233 also appears to be 
a very isolated system (MU08), implying that its disk structure should
be minimally affected by external perturbations.

\subsection{The Data\protect\label{data}}
To investigate the vertical structure of the \HI\ disk of IC~2233, MU08
extracted a series of \HI\  intensity profiles perpendicular to the disk,
spaced at regular intervals. These profiles were derived from an
\HI\ total intensity image with a spatial resolution of
$\sim16''$. The position angle for the galaxy was taken to be
172$^{\circ}$, and the adopted center of the galaxy, ($x,z$)=(0,0), 
was the kinematic center established by
MU08: $\alpha_{J2000}=08^{h}~13^{m}~58.9^{s}$, 
$\delta_{J2000}=+45^{\circ}~44'~$~\as{27}{0}. 
(Here $x$ is the distance along the galaxy
major axis, with values increasing toward 
the north, and $z$ is taken as the
distance above the midplane, with positive values to the east.) 
MU08 fitted a single Gaussian to each intensity
slice and plotted the displacement of the computed centroid from the
midplane ($z$=0) 
as a function of distance along the major axis. The resulting
``corrugation curve'' for the \HI\ data is reproduced in the upper
left panel of Figure~\ref{fig:corplots}.

The arrows on the upper left panel
of Figure~\ref{fig:corplots} indicate the  
extent of the stellar disk of
IC~2233, as determined from the 25 magnitude 
arcsec$^{-2}$ $B$-band isophote. It can be seen that 
near the edge of the stellar disk, the \HI\ layer
twists to form an ``integral sign'' warp.  On the side of the disk
where the \HI\ layer is more extended, the layer bends back
toward the midplane at larger radii. However, at small and
intermediate projected radii, regular 
undulations are clearly visible across the disk of the galaxy. 
Furthermore, it can be seen that a portion of the galaxy near $x=0$ 
is displaced
from the kinematically-defined
center-of-mass. One possible explanation is that 
this is linked with the presence of an inner bar.

To insure that our derived \HI\ corrugation curve is not significantly
impacted by optical depth effects,
we have compared \HI\ intensity profiles extracted from 
channel images with 16$''$ resolution (presented in MU08) with the corresponding ones made
with more uniform weighting of the raw visibilities (i.e., with a robustness
parameter ${\cal R}$ = $-1$), which leads to images with $12''$
resolution, albeit with rms noise about 40\% higher. We find that
the respective profiles and centroids agree to within the errors, although
the higher resolution images resolve out some of the more diffuse \HI\
component.  The peak \HI\ brightness temperatures are $\sim 130$~K for
the $12''$ images and $\sim 108$~K for the $16''$ images, respectively, near the
position of the kinematic center of IC~2233 and at the systemic
velocity.  Examination of the profiles shows that the \HI\ is optically
thin with $\tau\lsim0.02$.  This is consistent with the generally
excellent agreement between the \HI\ and \HA\ rotation curves (see Figure~13 of MU08).

To better constrain the nature and origin
of the corrugations in IC~2233, we have now extended our analysis of
its vertical disk structure using a number of additional data sets,
spanning far-ultraviolet (FUV) to  far-infrared (FIR) wavelengths
(Table~1). For each wavelength, we have computed a corrugation curve in a
manner analogous to that used for the \HI\ data. The results are
presented in electronic form in Table~2.
We have compared the
positions of background radio sources and foreground stars
with catalogued values from the NRAO VLA Sky Survey (Condon et
al. 1998) and the Tycho Reference Catalogue (H$\o$g et al. 2000), 
respectively.
With the exception of the 70$\mu$m and 160$\mu$m images (where no
foreground or background sources useful for registration were detected), we 
find that astrometric registration between images at different
wavelengths is accurate to
within a small fraction of a resolution element. Therefore,
we assume that astrometric uncertainties do not
impact our results unless explicitly noted.

Because of the low surface brightness nature of IC~2233, we have also
assumed that optical depth effects have a minimal impact on our
analysis at all available wavelengths. Above we noted that the \HI\
optical depth of the galaxy is
extremely low. At visible wavelengths the optical depth of IC~2233 also appears to be
minimal. MU08 estimated
the $B$-band extinction toward the center of the galaxy to be
$\sim$0.7 magnitudes, but this is likely to be an upper limit, as 
fits to the radial surface brightness profiles of IC~2233 show that the
light distribution is more strongly peaked than predicted by a simple
exponential disk model, and both the WIYN images published by
MU08 and the {\it Hubble Space Telescope} images published by Seth et al. (2005) reveal
that there are very few optically thick dust clumps in this galaxy and no true dust 
lane. Because of this, optical depth effects are expected to be
negligible in all of the {\it Spitzer} bands. The optical depth of
IC~2233 at UV wavelengths is more uncertain. However, 
better characterization of this will require detailed radiative transfer
modelling (see e.g., Matthews \& Wood 2001)
and is beyond the scope of the present paper.

\clearpage
%
\begin{deluxetable}{lccccl}
\tabletypesize{\tiny}
\tablewidth{0pc}
\tablenum{1}
\tablecaption{Data Sets used for Vertical Structure Analysis of
IC~2233}

\tablehead{
\colhead{Descriptor} & \colhead{Wavelength} & \colhead{Telescope} &
\colhead{FWHM PSF ($''$)} & \colhead{Reference} & \colhead{Dominant Components
traced} 
}

\startdata

NUV & 1350-1750\AA\ & {\it GALEX} & 4.6 & 1 & 
young-to-intermediate age stars; B-star photospheres\\

FUV & 1750-3000\AA\  & {\it GALEX} & 4.6 & 1 & young-to-intermediate age 
stars ($<10^{8}$ yr); B-star photospheres\\

$R$ & 5631-7156\AA\ & WIYN & 0.6 & 2 & young-to-intermediate age 
stars (up to $\sim10^{9}$~yr)\\

\HA$^{\rm a}$ & 6534-6606\AA\ & WIYN & 0.7 & 2 & ionized gas
associated with young ($<10^{7}$ yr), massive stars \\

3.6$\mu$m & 3.18-3.94$\mu$m &  {\it Spitzer} (IRAC) &1.7 &  3 
& old stellar populations; M-star photospheres\\

4.5$\mu$m &4.02-5.02$\mu$m  &  {\it Spitzer} (IRAC) &1.7 & 3 
& old stellar populations; M-star photospheres\\


8.0$\mu$m$^{\rm b}$ & 6.44-9.38$\mu$m &  {\it Spitzer} (IRAC) & 2.0 &
3 & PAHs (linked with spiral arms, molecular clouds)\\

24$\mu$m & 18.0-32.2$\mu$m  & {\it Spitzer} (MIPS) &6 &  3 & 
hot dust near sites of high-mass star formation\\

70$\mu$m & 50.0-111.0$\mu$m  &  {\it Spitzer} (MIPS) &18 &  3 
&  warm dust linked with star formation\\

160$\mu$m &  100.1-199.9$\mu$m &   {\it Spitzer} (MIPS) & 40 &3 
& cool, extended dust; infrared cirrus\\

\HI\ & 21~cm & VLA & 16 & 2  & neutral, atomic gas\\

\enddata


\tablenotetext{a}{Continuum-subtracted; includes a modest ($<10$\%) 
contribution from [\NII].}
\tablenotetext{b}{Data are uncorrected for the contribution from stellar
photospheric emission.}

\tablerefs{(1) Gil de Paz et al. 2007; (2) MU08; (3) L. D. Matthews \&
K. Wood, in prep. }

\end{deluxetable}
\clearpage

\subsection{Results: Multiwavelength Corrugation Curves\protect\label{results}}
Figure~\ref{fig:corplots} shows corrugation plots derived from the
datasets summarized in Table~1. Alongside each curve, a
representation
of IC~2233 at the corresponding wavelength
is shown. For reference, each of the
panels shows overplotted a polynomial fit to the \HI\ data
points lying within the boundaries of the stellar disk 
(i.e., excluding 
the outer disk regions where warping dominates the
vertical structure). Gaps in the corrugation
plots near $x\approx50''-70''$ at some wavelengths are due to 
contamination from a foreground star.

Figure~\ref{fig:corplots} illustrates that excursions from the midplane 
of up to $\sim5''$ ($\lsim$250~pc) are present in a variety of 
tracers in IC~2233. 
In the case of the mid-infrared (3.6$\mu$m and 4.5$\mu$m) 
continuum, these highest-amplitude 
excursions occur only at the edges of
the disk and can be attributed to the disk warp; however,
at all other wavelengths shown in Figure~\ref{fig:corplots}, 
undulations with amplitudes of up to
$\sim5''$ are present at low and intermediate projected radii
along the disk. None of these vertical undulations are as regular
as those observed in \HI; nonetheless, some interesting relationships
emerge between the patterns seen in various tracers.

In the southern portion of the galaxy (i.e., along negative
$x$ values), we see that tracers of young to intermediate-age 
stellar populations
(NUV, FUV, \HA, and $R$-band) 
bend away from the plane with the same sign and
comparable 
amplitude to that observed in \HI\ ($\Delta z\sim+5''$). For the 
NUV and FUV light, there is some hint of a 
slight phase lag with
respect to the \HI\ curve. This correlation is largely maintained in
the northern portion of the galaxy, to roughly $x\lsim 40''$.  In contrast, further north
($x\gsim 70''$), the vertical
structure of these
tracers diverges sharply from what is seen in \HI; here, 
excursions are 
observed toward positive $z$, while the \HI\ layer bends toward
negative $z$ values. Only beyond $x\gsim120''$ do the vertical
centroids again converge to values similar to
those observed in \HI. Visual
inspection of the UV, $R$-band, and \HA\ images
suggests that 
higher frequency undulations may also be present in these bands. These
are difficult to quantitatively characterize, but
may account for some of the scatter seen in these tracers. 

Turning toward tracers of the dust component in IC~2233, 
vertical undulations are also
clearly visible. The corrugation curve derived from the 
8$\mu$m light (believed to be dominated by PAH emission)
exhibits a pattern comparable in shape to the \HI\ data, but with
larger dispersion. 
Both the 24$\mu$m and the 70$\mu$m corrugation curves also show
close agreement with the \HI\ data over much of the disk. 
The exception is the region between $x\approx 40''-100''$ where the
24$\mu$m and 70$\mu$m emission diverge sharply from the \HI.   
In other galaxies
both the 8$\mu$m and 70$\mu$m emission have 
been found to have distributions similar to the 
\HI\ gas (e.g., Helou et al. 2001; Walter et al. 2007),
although the PAH emission tends to correlate more closely with the
molecular gas and cold dust, while the 70$\mu$m tends to be most prominent in
the highest \HI\ column density regions.

The vertical structure of the cool dust traced by the 160$\mu$m light 
shows  relatively little correlation with any other wavelength. Our
sensitivity is too limited to detect the outer $\sim$20\% of the
galaxy at this wavelength, and in addition, 
the low resolution of these data 
($40''$) makes the uncertainty in the 
location of the zero point comparable to the
amplitude of the vertical excursions. Nonetheless, the data 
are clearly inconsistent with a flat, planar
distribution for the cool dust.  On the northern side of the galaxy,
the 160$\mu$m emission appears qualitatively similar to the 70$\mu$m
emission, although the amplitude of the vertical excursions
appears larger. On the
southern side of the disk, the 160$\mu$m emission appears to be
anticorrelated with the \HI\ (and with all the other
tracers). Convolving and resampling the 70$\mu$m image to match the resolution and
pixel size of the 160$\mu$m data, we were unable to reproduce either
of these trends from resolution effects alone.

The divergent behavior between several of the \popI\ tracers
on the northern side of IC~2233 complicates the description of the
galaxy as a simple 
undulating plane. While projection effects due to warping or to the different
radial extents of various tracers may make some contribution to 
these differences, there is no {\it a priori}
reason to expect that these effects should be more severe on the northern side
of the galaxy. Therefore, the observed differences between the vertical
displacements of different disk components should
offer some clues on the origin of
the corrugation patterns and their possible link to the regulation of star
formation in the galaxy. For this reason, we now examine some of these differences in 
greater detail.

Visual inspection of the UV, $R$, and \HA\ images between $x\approx70''$-$120''$ 
reveals that
the vertical intensity profiles in these bands are
dominated by the light associated with
a smattering of compact \HII\ regions located several arcseconds above the
plane. The brightest of these (near $x\approx80''$) has counterparts at
24$\mu$m and 70$\mu$m. 
Interestingly, this  
position also corresponds to the location of the only major disparity between the \HI\  
and \HA\ rotation curves for the IC~2233; here, the \HA\ curve dips
toward the systemic velocity by $\sim$~35~\kms\ (see Figure~13 of
MU08). This difference between the \HI\ and \HA\ rotation curves  suggests
that either there is an absence of ionized gas at the tangent point 
along this line-of-sight, or else that the ionized material along this
direction has been kinematically
perturbed. The former scenario 
seems more likely given that we detect
significant \HI\ emission  near $x=80''$ at the same
velocity as the \HA\ emission (see Figure~11 of MU08).
However, the second possibility cannot be ruled out  (see also \S~\ref{tides}).

In the \HI\ total intensity image of IC~2233, we see a distinct \HI\ clump 
near $x\approx100''$ that is visibly
displaced toward negative $z$ values (Figure~1). Tracers of hot young stars (FUV,
NUV, \HA, $R$-band) all show a depressed surface brightness at
the corresponding location, although we do
detect counterparts to the \HI\ clump at
8.0$\mu$m, 24$\mu$m, 70$\mu$m, and 100$\mu$m. This dearth of starlight coupled
with the presence of FIR emission suggests that this sight-line may
intersect a young star-forming region where the gas is still predominantly
cold and dense. Recent studies show
that PAH emission tends to be a good proxy for dense
molecular gas in quiescent galaxies (e.g.,
Haas et al. 2002; Paladino et al. 2008), while emission at longer IR
wavelengths can be a hallmark of hot young stellar objects that are still highly
embedded. The disparity between the vertical structure of different
\popI\ tracers near this location  thus
could be partly a consequence of this region
being at an earlier evolutionary state compared with
some of the other optically prominent star-forming regions in the galaxy. 

\section{A Search for Velocity Corrugations}
Models for the origin
of structural corrugations in disk galaxies also predict corrugations in the
velocity field (see Alfaro et al. 2001 and references therein). 
The amplitude of the velocity corrugations and the 
relationship between velocity and structural 
corrugations can in principle be used to distinguish between
various models for their origin.
Unfortunately, with the exception of the Milky Way, it is 
difficult to measure both types of corrugations
in the same galaxy.

To explore whether any systematic patterns are discernible in the velocity
field of IC~2233, we have re-examined the observed \HI\ rotation curve 
from MU08. We have
parameterized the approaching and receding sides of the
curve with a 
Brandt (1960) function of the form:

\begin{equation}
V(x)=V_{\rm sys} + \left[V_{\rm max} 
\left(\frac{x}{x_{\rm max}}\right)\right]
\left[\frac{1}{3} + \frac{2}{3}\left(\frac{x}{x_{\rm max}}\right)^n\right]^{\frac{-3~}{2n}}
\end{equation}

\noindent where $V_{\rm sys}=554.7$~\kms, $V_{\rm
max}=$85.0~\kms, $x_{\rm max}$=\as{281}{4}, and $n$=1.09.
A comparison with this type of smooth, idealized curve 
can help to highlight subtle, systematic excursions from the
expected circular velocity at a given radius. In
Figure~\ref{fig:velres} we 
plot the residual velocity difference between
the observed rotation curve and the Brandt model as a function of distance
along the major axis. The dashed sections of the curve designate the
portions of the \HI\ disk where warping dominates the vertical structure
(see above). 

The residual curve in Figure~\ref{fig:velres} reveals that the
observed radial velocities along the IC~2233 disk exhibit
systematic positive and negative excursions from the Brandt
model, with amplitudes of $\lsim\pm$5~\kms. This type of systematic 
``waviness''
has  been seen previously in the rotation curves of other disk 
galaxies, particularly those with spiral
arms (e.g., de Vaucouleurs \& de Vaucouleurs 1963; 
Pi\c{s}mi\c{s} 1965; Fridman et al. 1998; Alfaro et al 2001). 
However, what is particularly striking in the case of IC~2233 is
that the locations of the local minima and maxima in the residual velocity
curve correspond closely
to the locations of the peak vertical structural
excursions observed in the \HI\ gas.

\section{Discussion\protect\label{discussion}}
As noted above, a number of mechanisms have been proposed to
account for the observed corrugations in disk galaxies.
We now discuss the suitability of several of these mechanisms for
explaining the observed corrugations in IC~2233.

\subsection{Tidal Interactions\protect\label{tides}}
Numerical modelling by Edelsohn \& Elmegreen  (1997) demonstrated that tidal
interactions with a low-mass 
companion are one means of producing a corrugated disk
structure. In this model, the
companion generates a spiral-shaped wave  that
propagates inwards from the edge of the disk, resulting in perpendicular
motions. Although the simulations of Edelsohn \& Elmegreen  
were performed for a  Milky
Way-like galaxy and a perturber roughly 8\% as massive, the results exhibit 
qualitative similarities to our observations of IC~2233. However,
MU08 found that IC~2233 appears to
be an extremely isolated galaxy, exhibiting no obvious signs of
a recent interaction or minor merger. This casts significant doubt upon
tidal effects as being the
primary driver of the vertical disk
structure of IC~2233. In
addition, since tidally induced corrugations are excited from the
outer regions of the disk (see Edelsohn \& Elmegreen 1997), this does
not provide a natural explanation for the ``null'' seen at the
kinematic center of IC~2233 (see 
\S~\ref{data} and Figure~\ref{fig:corplots}).
Interestingly, at least one of the galaxies
found by Florido et al. (1991) to have disk corrugations (NGC~4244),
also appears to be quite isolated (Dahlem et al. 2005).
A search for corrugations in additional isolated systems, coupled with
deep searches for low-mass companions, could provide
better statistical constraints on the likely role of companions in
exciting corrugated patterns. 

While an intrusion from a companion galaxy seems unlikely in the case
of IC~2233, the influence of impacts from very 
low-mass perturbers ($\lsim10^{6}~M_{\odot}$), such as
high-velocity clouds, cannot be entirely ruled out. 
Based on their energetics, high-velocity cloud impacts were the
explanation favored by Alfaro et al. (1991) to account for vertical
displacements of the midplane of the Milky Way and their associated
star formation activity (see also Franco et al. 1988).  Moreover, 
IC~2233 is known to have a vertically
extended \HI\ component that is kinematically distinct from
the main disk component (MU08), and whose presence suggests that  the
circulation of \HI\ material between the plane and higher
galactocentric latitudes is actively occurring (e.g., Fraternali \&
Binney 2008). While it would be difficult to explain how a {\it global}
corrugation structure as regular
as that seen in the \HI\ disk of IC~2233 could be explained by a series of cloud
impacts, such  events may help to account for {\it local} vertical displacements
and offsets between certain disk components. For example, such an event offers
one possible explanation for the differences in the vertical structure
of various disk tracers that is observed over a portion of the
northern half of IC~2233 as well as the perturbation in the \HA\
rotation curve observed near this position (see \S~\ref{results}).

\subsection{Galactic Bores}
The results presented 
in Figure~\ref{fig:velres} seem to exclude another class of model for
explaining the
vertical meanderings of the IC~2233 disk. In
particular, our findings appear to be inconsistent with a scenario where
corrugations result from gas being
hydraulically ``lifted''
off the plane by a spiral arm as a result of interaction of the spiral
density wave with a thick, magnetized,
gas disk (Martos \& Cox 1998). In this so-called ``galactic bore''
model, the lifting effect is expected to show an even symmetry about
midplane  (see Alfaro et al. 2001), resulting in some thickening of
the plane and
no net velocity excursions in an edge-on galaxy. 

\subsection{Parker Instabilities}
In contrast to the galactic bore picture, 
the undular mode of the Parker (1966) instability in
a thick, magnetized gas disk predicts a velocity pattern similar
to that seen in Figure~\ref{fig:velres} (see Franco et al. 2002). 
However, an additional prediction
of this model is that density enhancements should be present at
the locations of maximum deviation from the mean plane (e.g., Hanawa
1995). Indeed, such a trend 
has been observed in the corrugated Carina-Sagittarius arm of the Galaxy
(Franco et al. 2002). 
While IC~2233 does show a series of \HI\ density
enhancements along its disk (MU08), we find no clear correlation between
the locations of these clumps and the maxima in the corrugated pattern
in the \HI\ disk. 
Because corrugations resulting from the undular mode
of the Parker instability occur along the azimuthal direction,
line-of-sight integration effects may make it more
difficult to match the location of density enhancements
and vertical displacements 
in an edge-on galaxy. However, for the case of IC~2233, the Parker
instability  model
encounters an additional challenge in that
magnetic fields in small spiral galaxies with rotational velocities
comparable to IC~2233 are typically quite weak (e.g., 
Chy{\. z}y et al. 2007). 

\subsection{Spontaneous Gravitational Instabilities}
The observation that the corrugated structure of IC~2233 is more
pronounced in the younger stellar components than in the old stars 
would seem to argue against
a purely gravitational origin for the corrugations, since both gas and
stars will be influenced by gravity. However, gas is in general
more dynamically responsive than stars owing to its dissipative nature and 
its lower velocity dispersion (e.g., Kim \& Ostriker
2007). In addition, gaseous distortions may be amplified by
self-gravity and/or magnetic effects (e.g, 
Edelsohn \& Elmegreen 1997). For these reasons, even if the
corrugations have a purely gravitational origin, it is not entirely 
unexpected to find the largest amplitude corrugations in the gaseous
components  of the
galaxy. In this picture, 
the youngest stars (which still reside near their birth
locations) would be expected to 
exhibit a corrugated structure, while dynamical
evolution and heating would slowly erode these 
signatures in the older stars.

To determine whether an origin for the corrugations in
IC~2233 via gaseous instabilities 
is feasible, it is valuable to compare the ages of the stellar
populations participating in the corrugations with relevant
dynamical timescales in the disk.
For example, 
using equation 6-85 from Binney and Tremaine (1987), we can estimate
the oscillation period, $P$, for stars in the disk of
IC~2233. Assuming a  mass equal to the dynamical mass of the galaxy
($M_{d}=1.55\times10^{10}~M_{\odot}$), a disk radius of 9.2~kpc, and a
maximum rotational velocity $V_{\rm max}=85$~\kms\ (MU08), then at
intermediate galactocentric radii (near $r$=4.5~kpc),
we find $P\sim5\times10^{6}$~years. This is comparable to the
lifetime of an \HII\ region, implying
that the hot young stars giving rise to these ionized regions 
must be born near their present off-planar locations. 
If the corrugations arise initially in the gas, then the presence of
corrugations in the NUV and FUV light (corresponding to stars with
ages $<10^{8}$ years) implies that the midplane
deviations need to persist for only a few oscillation periods to be
visible in these bands. The mix of stars dominating 
the $R$-band light in IC~2233 is more
uncertain, although the very blue color of this galaxy ($B-R$=0.67;
MU08) suggests that this band may too be weighted toward 
moderately young stars, with ages no more than a few oscillation periods.

As discussed 
by Fridman \& Polyachenko (1984), vertical displacements in a galactic disk
may arise  from two types of internal gravitational 
mechanisms: spiral
density waves and vertical bending oscillations (see also Fridman et
al. 1998). Variations of both types of mechanisms
could be linked with disk corrugations, even in galaxies
with negligible magnetic fields. For example,
Nelson
(1976) suggested that corrugations could arise from the perturbations
of a spiral potential on a differentially rotating gas layer (see also
Nelson \& Matsuda 1980), while Masset
\& Tagger (1997) postulated that their origin could be the result of
a non-linear coupling between a spiral density wave 
and two so-called ``warp waves''. In the model of Masset \& Tagger, 
a transmitted wave results in an
integral sign warp, while a reflected wave produces
corrugations. 
Several arguments suggest that IC~2233 is likely to exhibit at least
rudimentary spiral structure (see MU08), but
unfortunately its edge-on orientation makes it impossible to
discern the details of this spiral pattern and thus to examine any possible
relationship to the observed corrugations.

Vertical bending oscillations in galaxy disks have been explored by
various authors as a possible origin for galaxy warps (e.g., Sellwood 1996;
Revaz \& Pfenniger 2004), but to our knowledge, relatively little attention
has been paid to these oscillations as a possible origin of disk
corrugations (although see Sparke 1995; 
Fridman et al. 1998; Griv \& Chiueh 1998). While the sophisticated
numerical simulations needed to explore this phenomenon are beyond
the scope of the present paper, we point to the results of some
existing studies that suggest that such an exploration might be
fruitful. For example, in certain galaxy disk models, 
Sellwood (1996) found higher order bending
modes whose amplitude and wavelength are reminiscent of
the corrugation pattern of IC~2233 (e.g., his Figure~1d), although it
remains to be explored how 
these results would translate to a disk containing both
gas and stars. More recently, Revaz
\& Pfenniger (2004) used multi-component disk models to show 
that bending instabilities can be 
long-lived (i.e., persisting through several rotations) 
provided that disks have sufficient self-gravity. In their
models it is assumed that some portion of the galaxy's dark matter
resides in the disk rather than in a halo. Revaz \& Pfenniger
explored only the low-order ($m=0$ and $m=1$) 
bending modes relevant to producing
galaxy warps; however, a search for higher-order modes, as well as an
extension of their models to include the dissipational behavior of the
gas would be of interest. 

\section{Summary}
We have presented a multi-wavelength analysis of the vertical
structure of the late-type, edge-on galaxy IC~2233. Systematic
excursions from the mean plane with amplitude $\lsim5''$
($\sim$250~pc) are observed in a variety of \popI\
tracers in this galaxy, including the neutral hydrogen gas, the young
stars, the \HII\ regions, and dust components of various
temperatures. This corrugated pattern appears most regular in the \HI\
gas, and is absent from the old stars traced by the mid-infrared
light. 

Over a significant fraction of IC~2233, the vertical
deviations observed in a variety of \popI\ tracers show similar signs
and amplitudes. A notable exception occurs along the disk region
between $x\approx 70''$-$120''$, thus complicating the interpretation
of the corrugation pattern in terms of a simple, global pattern. 
One contribution to  the differences seen along this region could be
variations in the filling
factors for various disk tracers.
Star-forming sites may also have different spectral
energy distributions depending on their ages, with the youngest being
dim at UV and optical wavelengths. A third possibility is that
perturbations (such as impacts from infalling gas clouds) may
locally influence the vertical structure of the disk.
Despite these complexities, we find that 
the locations of maximum deviation from the plane in the \HI\
gas correspond to perturbations in the \HI\ rotation curve with
amplitude $\lsim 5$~\kms.

A number of different scenarios have been suggested in the past
literature to account for the
corrugated structure in disk galaxies, and we have evaluated their
applicability to the case of IC~2233. Models invoking tidal
interactions with a companion, Parker instabilities, or a galactic bore all encounter
significant challenges in explaining the radial, galaxy-wide
structural and velocity 
corrugations of IC~2233. Models involving gravitational instabilities
of the gas layer appear
to be more promising. However, sophisticated numerical modelling will
be needed to evaluate such models further. The presence of a
corrugated structure in this rather low surface brightness galaxy
implies that vigorous star formation is not a necessary condition for
the appearance of complex vertical structure in the disks of galaxies.
Indeed, the quiescent nature of IC~2233 coupled with its low star
formation density is undoubtedly an important factor in allowing us to
observe the correspondence between vertical structures at a variety of
wavelengths despite our edge-on viewing angle.

\acknowledgements 
We thank the referee for thoughtful comments that helped to improve
this manuscript.
A portion of this worked was based on observations
from the {\it Spitzer Space Telescope}, operated by the Jet Propulsion
Laboratory (JPL), California Institute of Technology,
under contract 1407 with the National Aeronautics and Space
Administration (NASA). Financial support for this work was
provided to LDM through JPL Contract 1279242. The National Radio Astronomy Observatory
(NRAO) is a facility of the National Science Foundation, operated under cooperative
agreement by Associated Universities, Inc. The \HI\ data presented
here were obtained as part of the NRAO observing program AM649.

\clearpage

\begin{figure}
\vspace*{-7mm}
\centering
\includegraphics[width=0.7\textwidth]{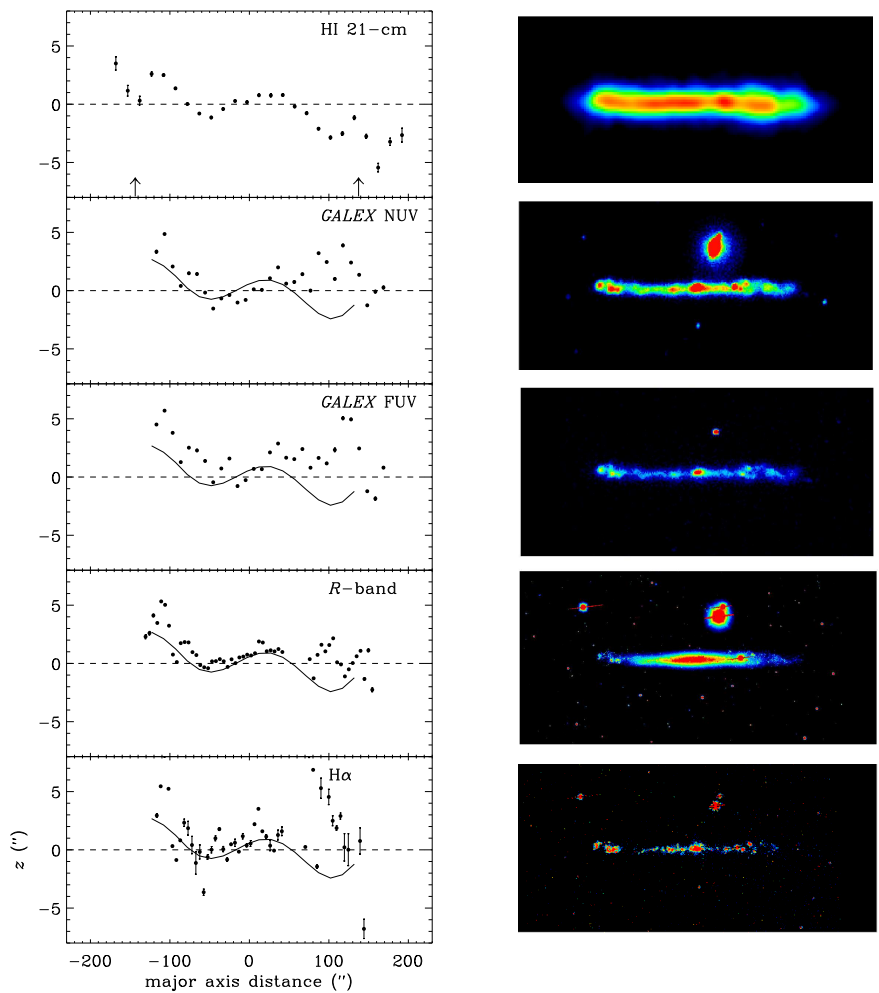}
\caption{Corrugation plots for IC~2233 derived at various
wavelengths, together with images of the galaxy in each band. 
The image panels are $210''\times450''$ across and are 
centered at $(x,z)=(0,0)$.
In the upper left panel, arrows denote the edges of the stellar
disk defined by the 25 mag arcsec$^{-2}$ $B$-band isophote. In
all other panels, the solid curve shows a polynomial fit to the \HI\
data points
interior to the arrows in the top left panel. At the distance of IC~2233,
$1''\approx$50~pc.}
\label{fig:corplots}

\end{figure}

\clearpage

\begin{figure}
\figurenum{1}
\vspace*{-7mm}
\centering
\includegraphics[width=.7\textwidth]{f1b.ps}
\caption{cont.}

\end{figure}

\clearpage

\begin{figure}
\centering
\scalebox{0.7}{\rotatebox{90}{\includegraphics{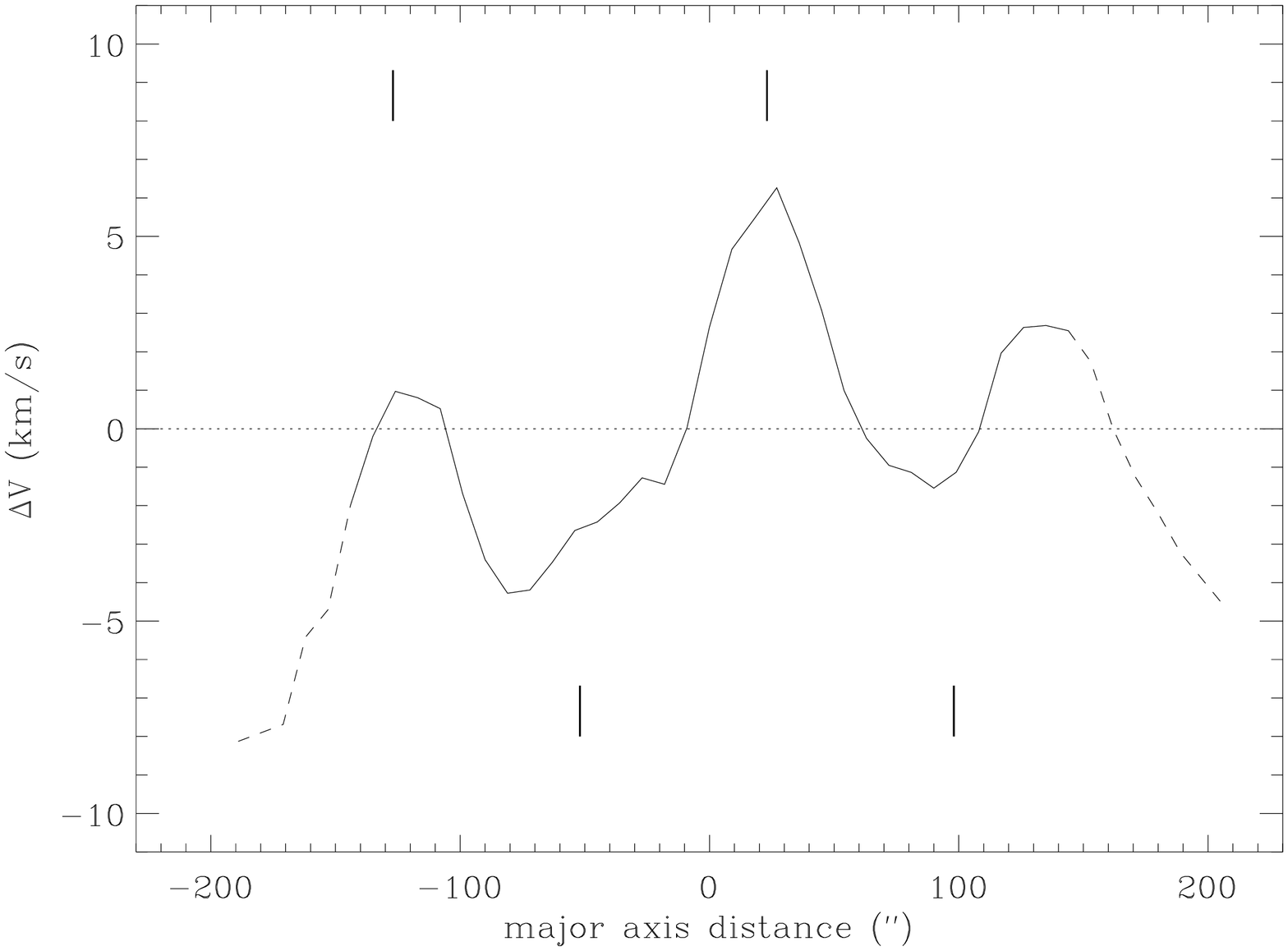}}}
\caption{Velocity residuals after subtraction of a Brandt
model rotation curve from the observed \HI\ rotation curve of IC~2233. The
residual curve has been smoothed to a resolution of $\sim30''$ using
a boxcar function. The dashed portions of the curve designate the
regions of the \HI\ disk where warping dominates the vertical
structure (see upper left panel of 
Figure~\ref{fig:corplots}). The hash marks indicate
the locations of the local minima and maxima of the vertical structural 
excursions of the \HI\ layer
as seen in Figure~\ref{fig:corplots}. }
\protect\label{fig:velres}
\end{figure}


\begin{deluxetable}{lccc}
\tabletypesize{\tiny}
\tablewidth{0pc}
\tablenum{2}
\tablecaption{IC~2233 Corrugation Curves}

\tablehead{\colhead{Band} & \colhead{Major axis distance ($x$)}
& \colhead{Vertical displacement ($z$)} & \colhead{$\sigma(z)$}\\
\colhead{} & \colhead{(arcsec)} & \colhead{(arcsec)} &
\colhead{(arcsec)} \\ 
\colhead{(1)} & \colhead{(2)} & \colhead{(3)} & \colhead{(4)} }

\startdata

NUV &  -106.72  &  4.86  &  0.04 \\
    & -96.52   & 2.07   & 0.05 \\
    &  -86.32  &  0.40  &  0.04\\
    & -76.12  &  1.49   & 0.08\\
    & -65.92  &  1.43   & 0.08\\
    & -55.72  & -0.17   & 0.06\\
    & -45.52 &  -1.54   & 0.06\\
    & -35.32 &  -0.67   & 0.07\\
    & -25.12 &  -0.38   & 0.07\\
   & -14.92  & -1.02    &0.06\\
    & -4.72  & -0.80    &0.04\\
    & 5.48  &  0.11   & 0.06\\
   & 15.68  &  0.07   & 0.07\\
   & 25.88  &  1.06   & 0.03\\
   & 36.08  &  1.99   & 0.03\\
   & 46.28  &  0.60   & 0.06\\
   & 56.48  &  0.74   & 0.09\\
   & 66.68  &  1.41   & 0.08\\
   & 76.88  &  0.00   & 0.09\\
   & 87.08  &  3.21   & 0.03\\
   & 97.28  &  2.46   & 0.05\\
  & 107.48  &  1.00   & 0.09\\
  & 117.68  &  3.88   & 0.10\\
  & 127.88  &  2.41   & 0.06\\
  & 138.08  &  1.36   & 0.06\\
  & 148.28  & -1.26   & 0.08\\
  & 158.48  & -0.09   & 0.11\\
  & 168.68  &  0.27   & 0.11\\ 

\enddata

\tablecomments{Table 2 is available in its entirety in the electronic
edition of the {\it Astrophysical Journal}. Explanation of columns:
col. (1) wavelength of the observation (see Table~1 for further
details); col. (2): displacement along the galaxy major axis in
arcseconds, with respect to the kinematic center (see \S~2.1);
col. (3) displacement of the fitted centroid of the vertical intensity profile;
(4) 1~$\sigma$ uncertainty in the position of the vertical centroid. }

\end{deluxetable}

\end{document}